# Mutual cooperation and tolerance to defection in the context of socialization: the theoretical model and experimental evidence.

T.S. Kozitsina [a,b], I.V. Kozitsin [c,a], I.S. Menshikov [a,b] [1]

[a] Moscow Institute of Physics and Technology (National Research University), Moscow, Russian Federation
[b] Federal Research Center Computer Science and Control, Russian Academy of Sciences, Moscow, Russian Federation
[c] V.A. Trapeznikov Institute of Control Sciences, Russian Academy of Sciences, Moscow, Russia

The study of the nature of human cooperation still contains gaps needing investigation. Previous findings reveal that socialization effectively promotes cooperation in the well-known Prisoner's dilemma (PD) game. However, theoretical concepts fail to describe high levels of cooperation (probability higher than 50%) that were observed empirically. In this paper, we derive a symmetrical quantal response equilibrium (QRE) in PD in Markov strategies and test it against experimental data. Our results indicate that for low levels of rationality, QRE manages to describe high cooperation. In contrast, for high rationality QRE converges to the Nash equilibrium and describes low-cooperation behavior of participants. In the area of middle rationality, QRE matches the curve that represents the set of Nash equilibrium in Markov strategies. Further, we find that QRE serves as a dividing line between behavior before and after socialization, according to the experimental data. Finally, we successfully highlight the theoretically-predicted intersection of the set of Nash equilibrium in Markov strategies and the QRE curve.

---

[1] Deceased

# 1. Introduction

Human behavior is still a question and still contains gaps needing investigation. What we know is that our way of thinking, actions, and beliefs depend on many different factors: internal and external. Human behavior includes the important social ability of cooperation, defined by the Cambridge dictionary as "the act of working together with someone or doing what they ask you"[1]. Perhaps more importantly, cooperation is about sharing mutual profit, equality, costs, and skills. Nowadays, during the pandemic, we realize how important cooperation is for society. It is not about money, but human lives. Examples of cooperation include wearing a mask, keeping social distance, and being patient and generous with public. Thus, studying people's ability to cooperate helps in making beneficial choices during the world pandemic[2].

Despite the evidence of the clear advantages of cooperation, rational to choose defection rather than cooperation when faced with a social dilemma. It is for this reason that these situations are called dilemmas, and studying the factors that lead to cooperation is an important step toward understanding this example of behavioral economics.

Authors make different arguments on which factors may increase cooperation in social dilemmas[3]: using communication[4,5] or socialization[6–9], mobility and dynamics[10,11], connectivity[12], or aspects of an individual's identity[13,14]. The choice to cooperate is more of an intuitive act than a meaningful one. It is an emotional, quick, automatic operation that does not involve effort. To support this claim, the authors compared the amount of time that participants in the experiments spent choosing between cooperation and non-cooperation strategies[15]. Their results indicate that quick choice could be a predictor of cooperation. Effects of sociality could also lead to increasing of cooperation[16]. Valerio Capraro even introduces the cooperative equilibrium for explaining deviation from Nash equilibrium, based on the idea that people have some tendency to cooperate by default[17,18].

There are different approaches to shifting strategies from individual to social. The question remains regarding what models can explain irrational cooperation in social dilemmas. Here we list some concepts accepting relatively high cooperation level:

- Quantal response equilibrium[19]
- Level-k[20]
- Cognitive hierarchy[21]
- Quantal level-k[22]
- Trembling hand perfect equilibrium[23]
- Proper equilibrium[24]

Previous studies demonstrate that social interaction significantly increases the cooperation level in iterated Prisoner's Dilemma games, from a 20% cooperation rate prior to socialization to 53% after socialization[9,25–27]. To model such a high level of cooperative strategy choice we require a specific approach. In the paper[26] it was proposed to consider Prisoner's Dilemma in Markov strategies. For this game, a symmetric totally mixed Nash equilibrium was found. However, this equilibrium better fits strategies prior to socialization than after. Therefore, we developed a new model that is able to describe high-cooperation strategies.

## 2. Model

### *2.1. Prisoner's Dilemma game (PD)*

This work is based on the broadly-known Prisoner's Dilemma game. In this game, two participants choose between two strategies: Left or Right for the first strategy, and High and Low for the second. The choices are simultaneous and independent from each other. Payoffs correspond to the following payoff matrix (see Table 1), which were employed in the model and experiments. Strategies have the next sense: Left and High for Cooperation, Right and Low for Defection.

Table 1. Payoff matrix in Prisoner's Dilemma game.

| Payoff | Left | Right |
|---|---|---|
| **High** | 5, 5 | 0, 10 |
| **Low** | 10, 0 | 1, 1 |

### *2.2. Nash equilibrium for PD*

PD has one Nash equilibrium: it is a mutual choice of Defect strategy which gives the payoff of 1 for two players. However, laboratory experiments show that people in some conditions avoid Nash equilibrium[9,26]. For example, under social framing, individuals may start to choose more frequently the Cooperate strategy, a sort of behavior that could be considered irrational. For this reason, it would be interesting to discern a theoretical concept underlying this specific behavior.

### *2.3. Nash equilibrium for PD in Markov strategies*

The papers[7,9,28,29] argue that for some subjects, social context led to the increase in cooperative choices up to 100%. So, the behavior under social context is far from Nash equilibrium. One of the ways to somehow describe the cooperative behavior is to consider Prisoner's Dilemma game in Markov strategies.

Consider two participants $i \in \{1,2\}$. Let us denote the probability to cooperate in round $t$ for the first participant as $p_1^c(t)$. We describe participants' behavior by means of the following two quantities: (1) $\gamma$ – *mutual cooperation* (probability of cooperative choice as the respond to cooperative choice of opponent on the previous round); (2) $\alpha$ – *tolerance to defection* (probability of cooperative choice as the response to a defective choice of opponent on the previous round). These two variables imply that individuals' strategies at round $t-1$ determine completely their behavior at round $t$. This model will be referred as PD in Markov strategies[26,30], and for brevity we will refer to subsequent $\gamma$ and $\alpha$ as Markov strategies. Dynamics of participants' actions can be presented as follows:

$$\begin{cases} p_1^c(t) = \gamma_1 p_2^c(t-1) + \alpha_1\big(1 - p_2^c(t-1)\big), \\ p_2^c(t) = \gamma_2 p_1^c(t-1) + \alpha_2\big(1 - p_1^c(t-1)\big). \end{cases} \quad (1)$$

In a stationary state, we have:

$$\begin{cases} p_1^c = \gamma_1 p_2^c + \alpha_1(1 - p_2^c), \\ p_2^c = \gamma_2 p_1^c + \alpha_2(1 - p_1^c), \end{cases} \quad (2)$$

where $p_1^c$ and $p_2^c$ are stationary probabilities of cooperation.

Payoff function for participant 1 has the following form:

$$U(p_1^c, p_2^c) = -4p_1^c p_2^c - p_1^c + 9p_2^c + 1. \quad (3)$$

The paper[26] found a symmetric (whereby $\gamma_1 = \gamma_2 = \gamma$ and $\alpha_1 = \alpha_2 = \alpha$) totally mixed Nash equilibrium for Prisoner's Dilemma in Markov strategies in explicit form. This equilibrium can be represented as the points $(\alpha, \gamma)$ that meet equation

$$5\alpha^2 + 9\gamma^2 - 14\alpha\gamma - 10\gamma + 1 = 0 \quad (4)$$

and located in the unit square (see Figure 1). Further we will refer to equilibrium as Nash equilibrium and this curve as Nash equilibrium curve.

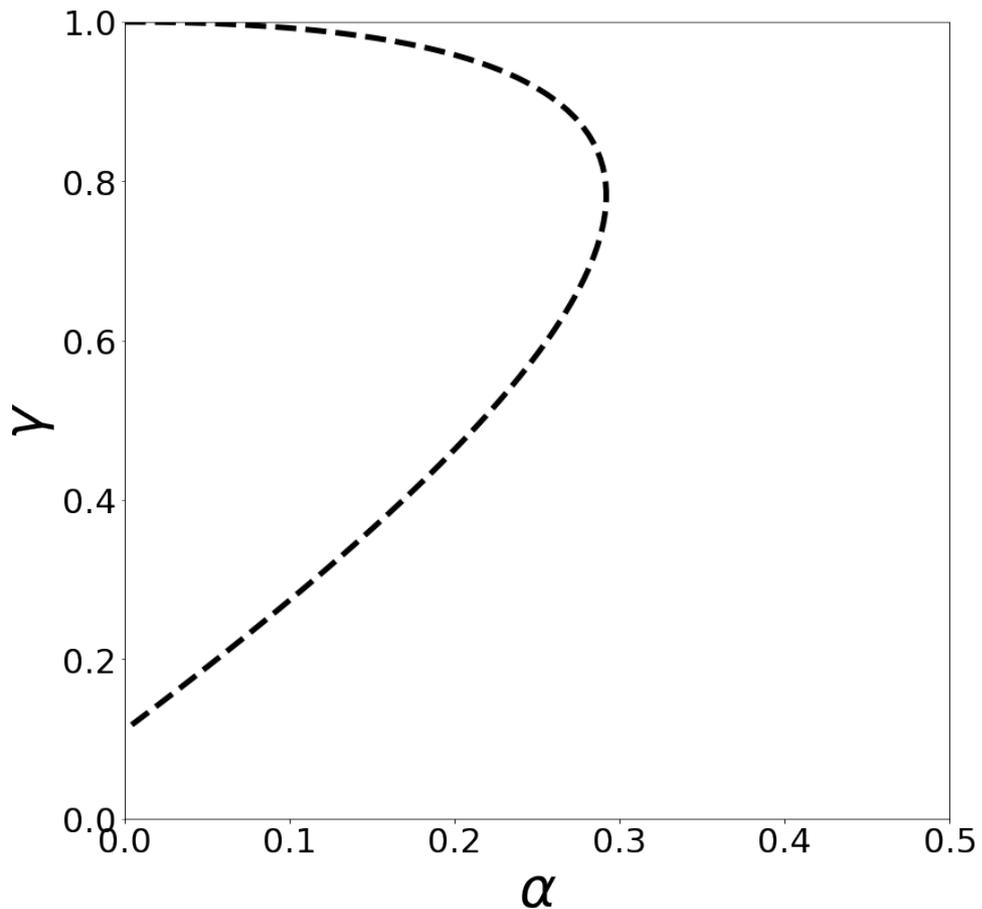

*Figure 1.* Symmetric totally mixed Nash equilibrium for PD in Markov strategies

It is evident from Figure 1 that curve (4) exists in the area which is characterized by relatively small values of $\alpha$ (the tolerance to defection does not exceed 0.3). However, experimental results (Section 4) show that tolerance to defection could be even more than 0.5. Attempting to reconcile this problem, we derive QRE for PD in Markov strategies in the next Section.

## 3. Quantal response equilibrium (QRE)

The QRE model was invented to explain the observed behavior of participants in laboratory experiments when it differs significantly from the Nash equilibrium[19]. QRE is an internally consistent equilibrium model in the sense that the quantum response functions are based on the distribution of equilibrium probabilities in the choice of strategies of opponents, not simply on arbitrary beliefs that players may have about these probabilities. One of the features of the model is that it allows consideration of "players making mistakes". QRE imposes a requirement that expectations must match an equilibrium choice of probabilities. However, in contrast to the classical Nash equilibrium, the definition of QRE assumes that participants strive for the best answer only in the probabilistic sense: the better the answer, the more likely the participant will choose it[31,32]. The QRE was compared with experimental data, and determined that this approach provided better fit than the Nash equilibrium[33]. In practice, the QRE is built upon employing logistic distribution. Answer $s_i$ to the mixed strategy $s_{-i}$ of the remaining players (the probability of choosing strategy $s_i$) is expressed through the following formula:

$$\sigma_i(s_i|s_{-i}) = \frac{e^{\lambda U_i(s_i, s_{-i})}}{\sum_{s_i'} e^{\lambda U_i(s_i', s_{-i})}}, \qquad (5)$$

where $\lambda$ – is the parameter of participant's rationality, and $U_i(s_i, s_{-i})$ is the expected gain of participant $i$ when strategies of other players $s_{-i}$ and strategy $s_i$ of participant $i$ are given. Therefore, when $\lambda \to 0$ (low rationality) choices are equally random, and when $\lambda \to \infty$ (high rationality) participants chose the strategies with the highest expected payoff.

The paper[26] demonstrated that the concept of QRE for simple PD game works only when the probabilities of cooperative choices do not exceed 50%. Studying high cooperation, we found QRE for PD in Markov strategies. Consider $\{\alpha_1, \gamma_1\}$ – Markov strategies of player 1, and $\{\alpha_2, \gamma_2\}$ – Markov strategies of player 2, $\{\lambda_1, \lambda_2\}$ – players' rationalities. We found QRE for this game by fixing both parameters of one of the players (for example, player 2's parameters $\alpha_2$ $and$ $\gamma_2$) and only one parameter of another player (for example, $\gamma_1$). We observed the remaining parameter under two conditions (for example, $\alpha_1 = 0$ and $\alpha_1 = 1$). It provides the combinations of specific strategies of players 1 and 2. Following this, we switched to the symmetrical case and used $\alpha_1 = \alpha_2, \gamma_1 = \gamma_2, \lambda_1 = \lambda_2$. Finally, the following system provided us QRE strategies for PD in Markov strategies (the detailed description of QRE finding can be found in Appendix 1):

$$\begin{cases} \alpha = \dfrac{e^{\lambda U|_{\alpha=1}}}{e^{\lambda U|_{\alpha=0}} + e^{\lambda U|_{\alpha=1}}}, \\ \gamma = \dfrac{e^{\lambda U|_{\gamma=1}}}{e^{\lambda U|_{\gamma=0}} + e^{\lambda U|_{\gamma=1}}}, \end{cases} \quad (6)$$

where $\alpha \in [0; 1], \gamma \in [0; 1]$, $U$ is the payoff function (expressions for $U|_{\alpha=0}, U|_{\alpha=1}, U|_{\gamma=0}$, and $U|_{\gamma=1}$ are given in Appendix 1), and $\lambda \in [0; +\infty)$ is fixed. We propose to solve system (6) numerically, reducing it to finding (as far as feasible) the optimal solution of the following optimization problem:

$$\begin{cases} \min_{\alpha,\gamma} (\dfrac{e^{\lambda U|_{\alpha=1}}}{e^{\lambda U|_{\alpha=0}} + e^{\lambda U|_{\alpha=1}}} - \alpha)^2 + (\dfrac{e^{\lambda U|_{\gamma=1}}}{e^{\lambda U|_{\gamma=0}} + e^{\lambda U|_{\gamma=1}}} - \gamma)^2, \\ \alpha \in [0; 1], \\ \gamma \in [0; 1]. \end{cases} \quad (7)$$

Let us first investigate the behavior of the objective function

$$f = (\dfrac{e^{\lambda U|_{\alpha=1}}}{e^{\lambda U|_{\alpha=0}} + e^{\lambda U|_{\alpha=1}}} - \alpha)^2 + (\dfrac{e^{\lambda U|_{\gamma=1}}}{e^{\lambda U|_{\gamma=0}} + e^{\lambda U|_{\gamma=1}}} - \gamma)^2. \quad (8)$$

In Fig. 2, we demonstrate how solutions obtained correspond to the contour lines of the objective function for different values of rationality. We observe that for values of $\lambda$ close to null, the (unique) minimum of the objective function is reached near the point [0.5, 0.5] that corresponds to the sense of $\lambda$ (under assumption of low rationality, individuals should act at random). With the increasing of $\lambda$, we notice several local minima gradually shifting to the Nash equilibrium curve. When rationality is high, local minima converge to the Nash equilibrium curve on one hand and the strategies' profile corresponding to the standard Nash equilibrium $\alpha = 0, \gamma = 0$ (defect/defect) on other hand. This result corresponds to the theory, as the Nash equilibrium describes behavior of fully rational actors[19].

To solve optimization problem (7), we use the Python package *minimize* from scipy.optimize[34]. Fig. 3 plots the arrangement of obtained a symmetrical quantal response equilibrium for PD in Markov strategies which form a near smooth curve in the range of small $\lambda$ (approximately less than 5). For these values of rationality, the objective function always has a unique global minimum, which is perfectly caught by the solver. In the middle range of $\lambda$ (approximately in the interval $[5, 7.08]$), the solution of the optimization problem approaches the Nash equilibrium curve. Nonetheless, at these levels of rationality, the QRE curve loses its smoothness and solutions "leapfrog" on the Nash equilibrium curve (see blue triangles on Fig. 3). For large values of the rationality ($\lambda > 7.08$), solutions of (7) converge to the point $\alpha = 0, \gamma = 0$ which does not belong to Nash equilibrium curve in Markov strategies. Instead, it marks the strategies' profile of standard Nash equilibrium (defect/defect). According to the theory, intersections of the Nash equilibrium curve and the QRE curve should exist and mark the branch of Nash equilibrium in Markov strategies that appear to have a special significance in experimental data description[19]. We demonstrate that there are few intersections of these curves (blue triangles, Fig. 3) that could be the result of the optimization method weakness. However, exactly the "first" intersection (which is located near $\alpha \approx 0.2, \gamma \approx 0.5$ and is derived under $\lambda \approx 4$) fits the experimental data best when compared to other intersections (Section 4).

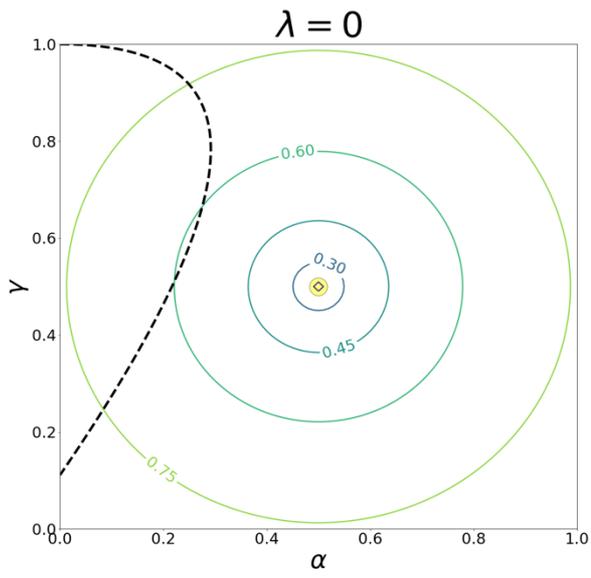
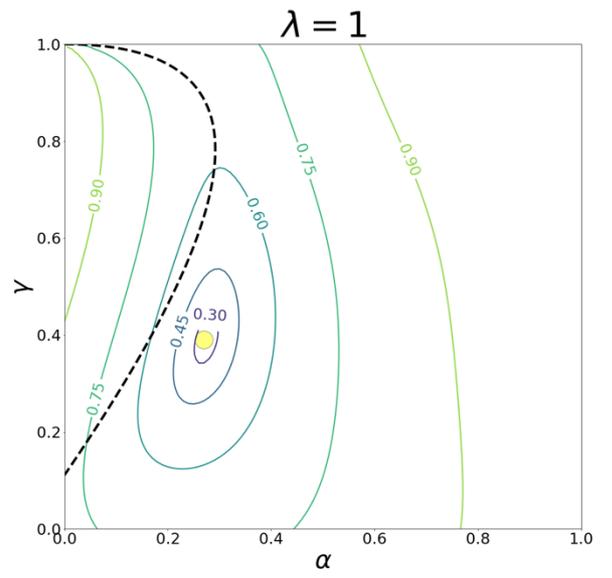
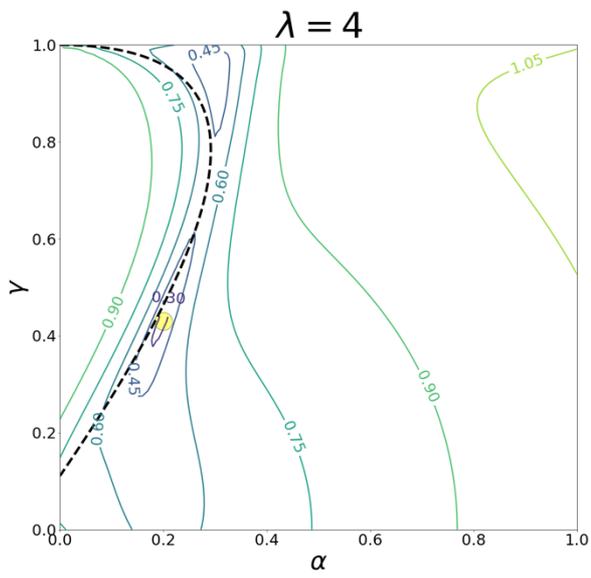
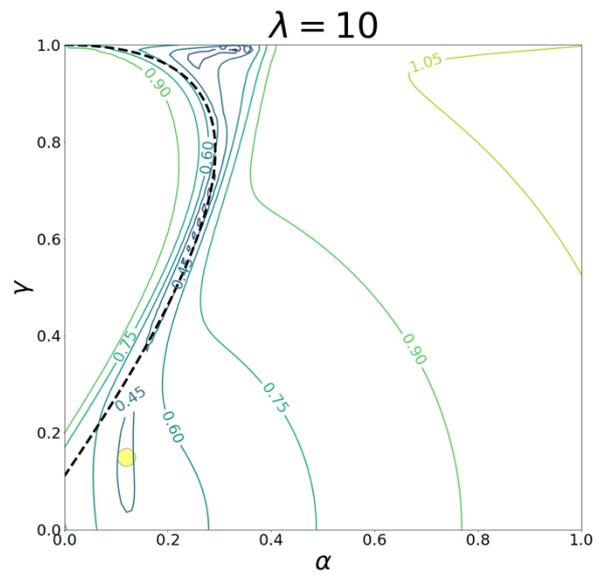
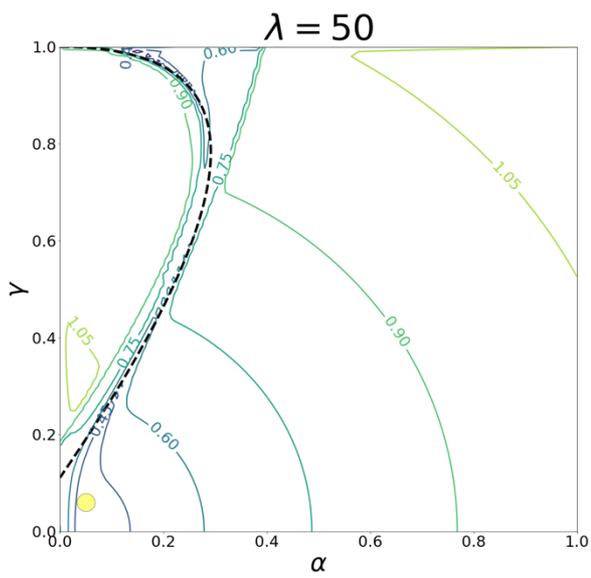
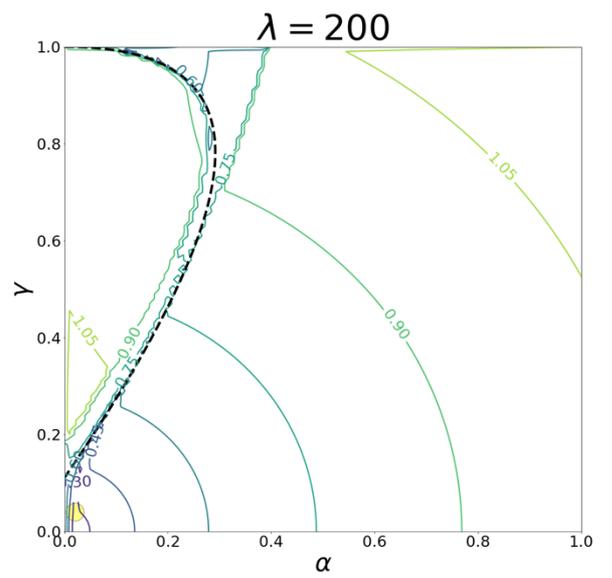

*Figure 2*. The panels plot contour lines of the objective function for different values of rationality. The dashed line represents Nash equilibrium for PD in Markov strategies. Yellow circles represent solutions of the corresponding optimization problem.

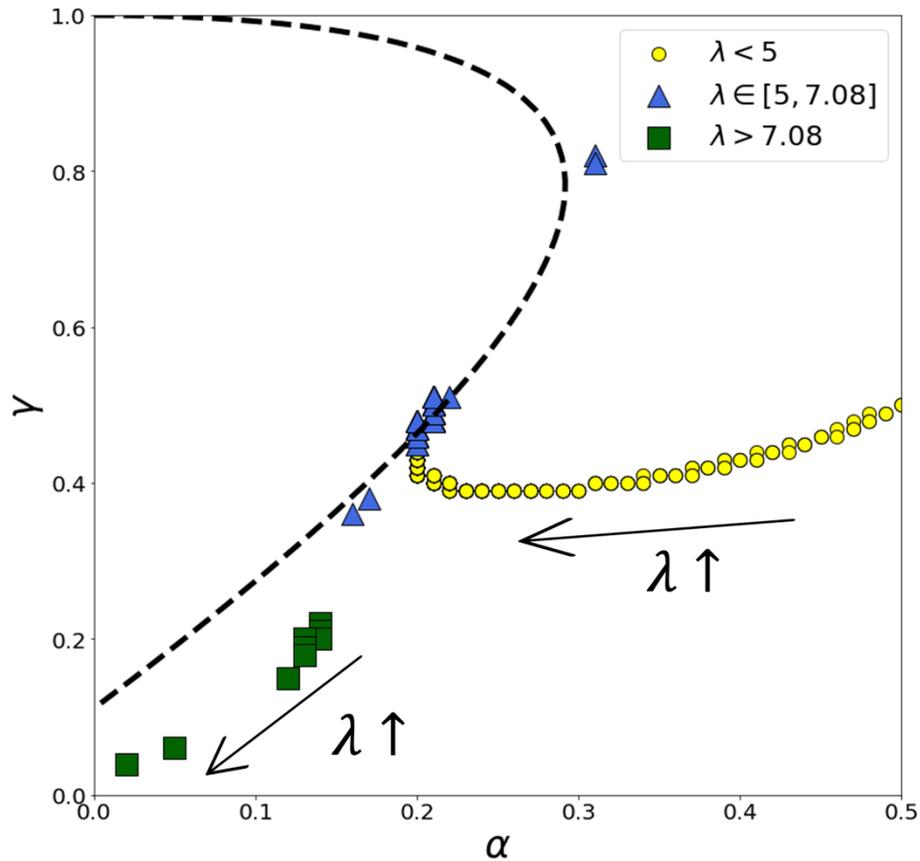

*Figure 3*. QRE for PD in Markov strategies derived using solving optimization problem (6) for different values of rationality. The arrows indicate the direction in which rationality grows. The resulting QRE curve consists of three branches that correspond to different ranges of rationality. The dashed line represents Nash equilibrium for PD in Markov strategies.

## 4. Experimental results and discussion

In this section, we compare the equilibrium found against the data from laboratory experiments which were presented in the following publications[9,25,26]. The general goal of these experiments was to identify the effect of socialization on the level of cooperation choice in the PD game.

The full description of the experiments can be found in Appendix 2. The following is a schematic representation of the experimental design:

1. 12 recruited participants (all strangers).

2. Participants play iterated PD (Table 1) in a mixed-gender group of 12 people for 11-22 rounds.

3. Socialization of unacquainted members of groups. Division of participants into two groups of 6 people.

4. Participants play iterated PD (Table 1) in the newly formed groups for 15-20 rounds.

5. Participants are compensated for the experiment.

In Table 1 (Appendix 3) we present aggregated results of the experiments. We find that the choice of cooperation is higher after socialization (58%) rather than before (22%). We assume that socialization compensates for the irrationality of these choices. This implies that despite the expectation that payoff of defection is higher than of cooperation, the utility of sociality is higher than probable losses of the cooperation choice. In comparing theoretical results with the experimental data, we found for every part of the experiments probabilities of mutual cooperation (gamma) and tolerance to defection (alpha) (see Table 1, Appendix 3).

We first analyzed how experimental points correspond to values of objective function (8) under different levels of rationality (see Fig. 4). We observed that most participants' strategies can be approximated by the minima of the objective function after selecting the appropriate level of rationality. More precisely, we recognized that behavior of individuals with high level of cooperation (more than 50%) could be modeled by selecting low rationality rates (which was one of our objectives) whereas low-cooperative participants were well - approximated by high values of rationality. From this perspective, we conclude that socialization reduces the level of rationality. Unfortunately, participants located in the upper right zone of the phase plane are still unexplained. The fact that the QRE curve have intersections with the Nash equilibrium curve means that our results are consistent with the theory. Notably, among all Nash equilibrium in Markov strategies, the most "successful" equilibrium (in fitting experimental data) is one found on the "first" intersection between the QRE curve and Nash equilibrium curve (at the rationality level $\lambda \approx 4$). Interestingly, the QRE equilibrium for $\lambda\sim\ <4$ divides strategies before and after socialization (see Fig. 5), resembles phase boundary.

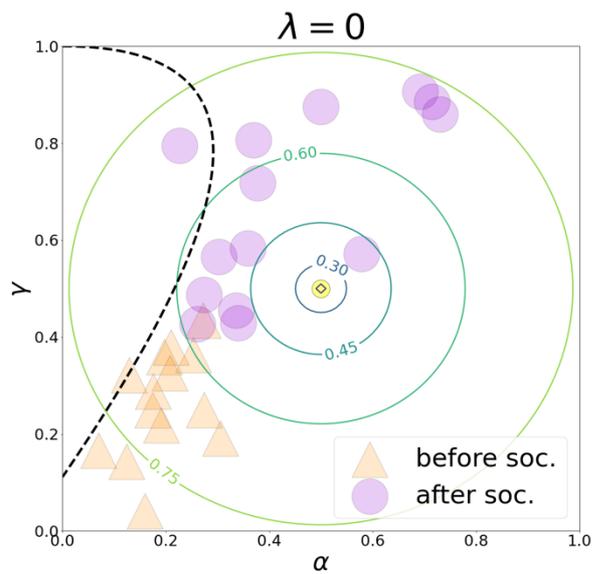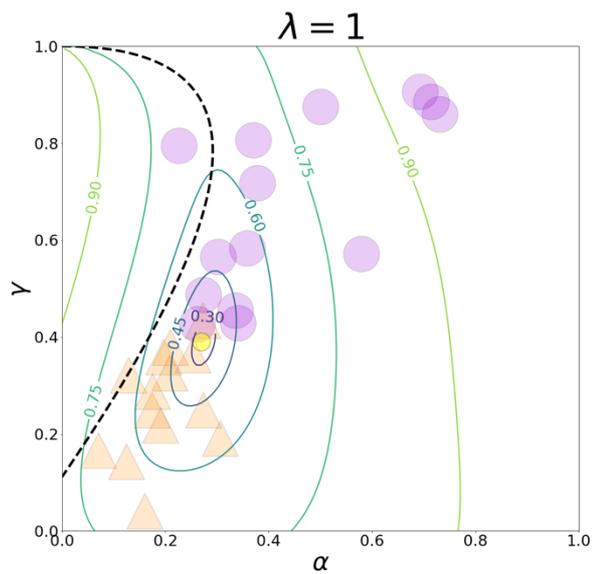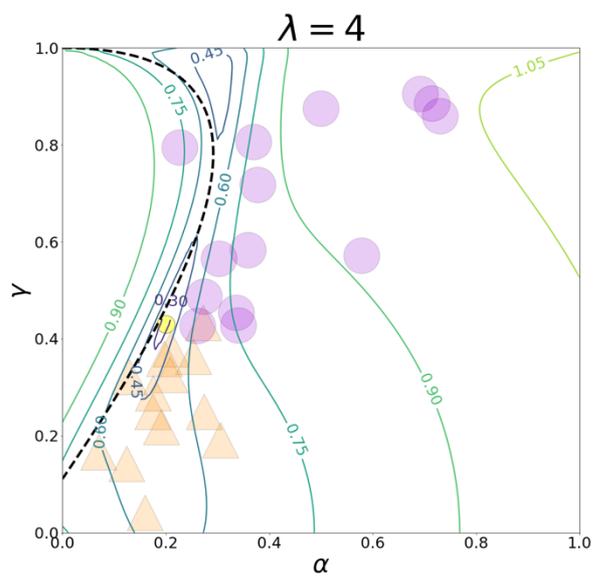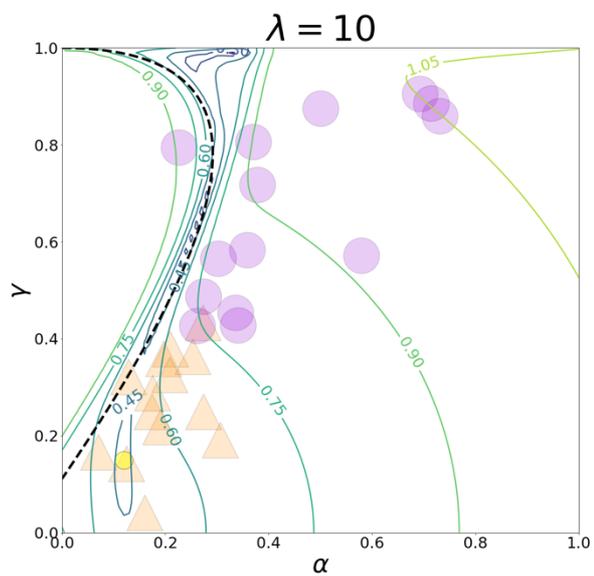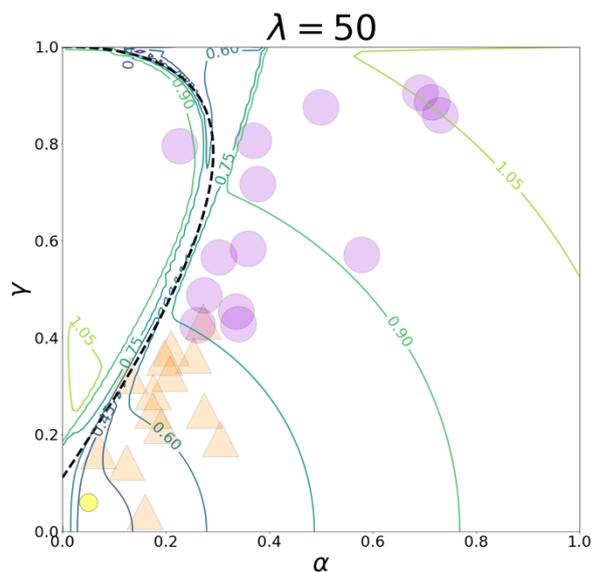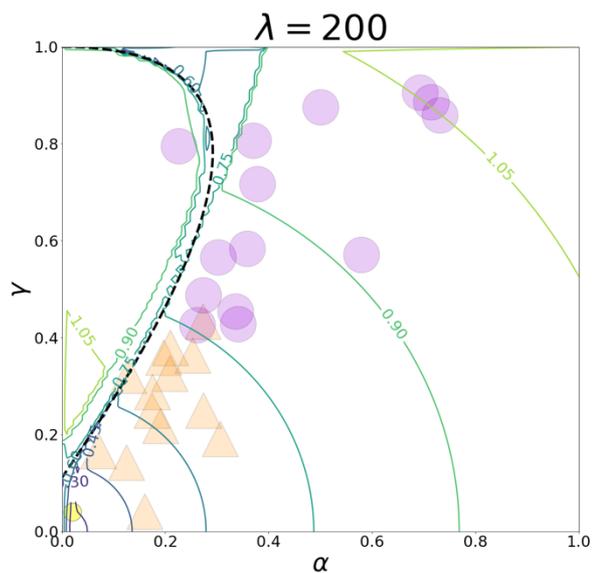

*Figure 4.* The panels plot contour lines of the objective function for different values of rationality. The dashed line represents Nash equilibrium for PD in Markov strategies. Yellow circles represent solutions of the corresponding optimization problems (QRE). Orange triangles indicate strategies of experiments' participants before socialization while violet circles represent strategies after socialization.

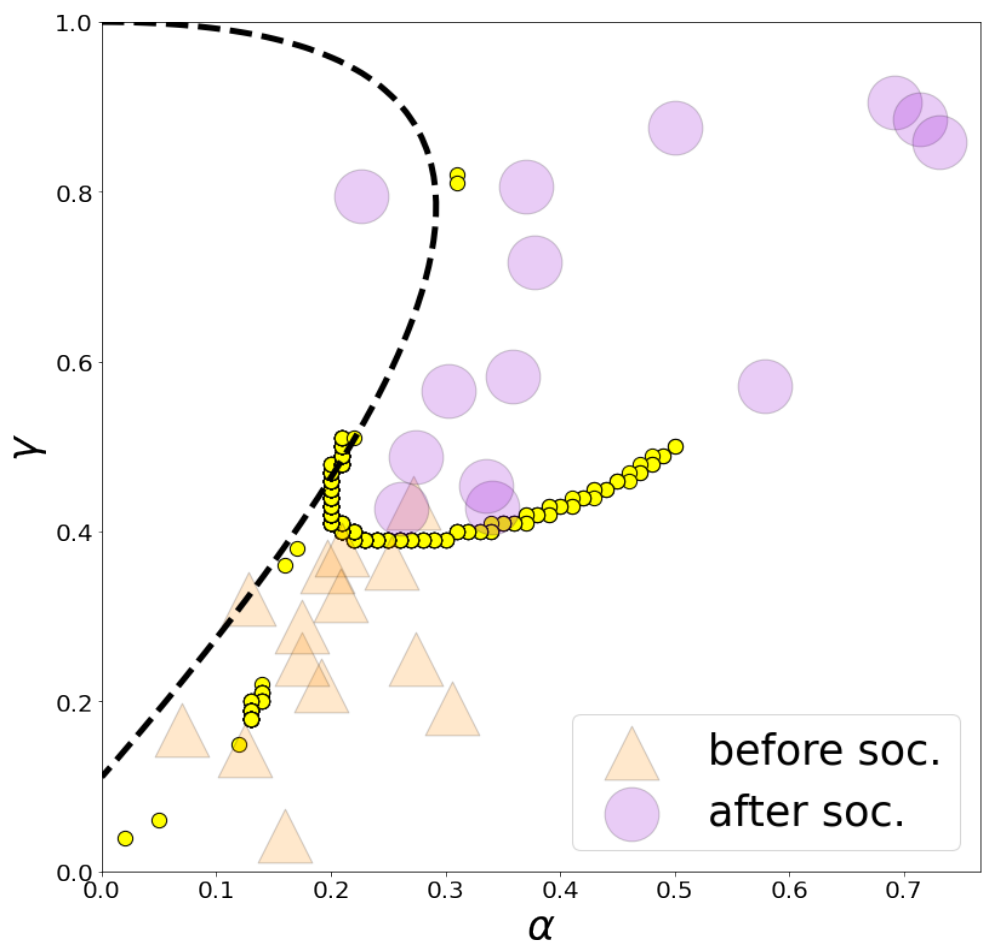

*Figure 5.* The dashed line represents Nash equilibrium for PD in Markov strategies. Yellow circles represent the QRE curve. Orange triangles signify strategies of experiments' participants before socialization, while violet circles signify strategies after socialization. QRE points for $\lambda \sim\, < 4$ serve as a natural border between points before and after socialization.

## 5. Conclusion

In the era of highlighting the importance of every individual's choice while mass media promotes living for oneself, it is crucial to remember cooperation as an effective mechanism to promote well-being of the whole society. Our paper proposes a new theoretical concept to explain the high levels of cooperation (more than 50%) which was previously obtained in the laboratory experiments[9,26] (examining the influence of socialization on strategies choices in the Prisoner's Dilemma game). We found a symmetrical quantal response equilibrium for Prisoner's Dilemma game in Markov strategies (QRE) and compared it with the symmetrical Nash equilibrium for this game[26] and experimental results[9]. Under the Prisoner's Dilemma game in Markov strategies, we specify the following: a) instead of pure PD strategies (cooperate or defect), we employ mutual cooperation (the probability of cooperative choice in response to an opponent's in the previous round) and tolerance to defection (the probability of cooperative choice as the response to an opponent's defective choice in the previous round); b) the choice of strategy in the current period depends solely on the strategies from the previous period.

When we matched QRE and experimental data, we observed that high level of cooperation after socialization can be explained by QRE with the low rationality rates. Conversely, low-cooperative results before socialization are well-approximated by high values of rationality. We also found how QRE completes the existing Nash equilibrium for this game. The intersection between the equilibrium curves under the low parameter of rationality ($\lambda \approx 4$) gives the unique selection of Nash equilibrium, fitting the experimental data the closest (comparing to other Nash equilibrium in Markov strategies). Additionally, QRE curve (for parameter of rationality $\sim < 4$) serves somewhat as a phase boundary for the experimental data before and after socialization: before socialization points lie below QRE while after socialization points lie above. However, we understand the chance this result was found by coincidence. Therefore, one possible continuation of the study is to investigate the possible property of QRE as a phase boundary in two directions: (1) theoretical (by deriving corresponding models) and (2) empirical (by conducting experiments). Finally, we also observed experimental strategies which deviate far from QRE or Nash equilibrium. This indicates that the investigation of theoretical concepts explaining the behavior of high cooperation is still in progress.

## 6. Acknowledgements

The authors are grateful to reviewers for their invaluable comments.

This research was supported by the grant in RFBR 19-01-00296A.

This paper is dedicated to the memory of I.S. Menshikov.

# Appendix 1

Here we find a symmetrical quantal response equilibrium (QRE) for PD in Markov strategies.

Consider $\{\alpha_1, \gamma_1\}$ – Markov strategies for player 1, and $\{\alpha_2, \gamma_2\}$ – Markov strategies for player 2, $\{\lambda_1, \lambda_2\}$ – parameters of rationalities for players 1 and 2 (rationalities are assumed to be constants).

We find QRE for this game using the following approach: fix both parameters of one player (for example, $\alpha_2$ and $\gamma_2$) and only one parameter for another player (for example, $\gamma_1$). After, we switch to the symmetrical case by assuming $\alpha_1 = \alpha_2, \gamma_1 = \gamma_2, \lambda_1 = \lambda_2$. The combination of pure strategies for another parameter then gives us the required equation of QRE. We must determine the equations of payoff function for different cases of strategy fixation.

Using system (2), we have:

$$p_1^c = \frac{\alpha_1 - \alpha_2(\alpha_1 - \gamma_1)}{1 - (\alpha_1 - \gamma_1)(\alpha_2 - \gamma_2)}, \quad p_2^c = \frac{\alpha_2 - \alpha_1(\alpha_2 - \gamma_2)}{1 - (\alpha_1 - \gamma_1)(\alpha_2 - \gamma_2)}$$

First, we fix the profile of strategies $\alpha_2, \gamma_2, \gamma_1$ and calculate the probabilities of cooperative choice for pure strategies $\alpha_1 = 0$ and $\alpha_1 = 1$.

For $\alpha_1 = 0$ we have:

$$p_1^c = \frac{\alpha_2 \gamma_1}{1 + \gamma_1(\alpha_2 - \gamma_2)},$$

$$p_2^c = \frac{\alpha_2}{1 + \gamma_1(\alpha_2 - \gamma_2)}.$$

For $\alpha_1 = 1$ we have:

$$p_1^c = \frac{1 - \alpha_2 + \alpha_2 \gamma_1}{1 + (1 - \gamma_1)(\alpha_2 - \gamma_2)},$$

$$p_2^c = \frac{\gamma_2}{1 - (\alpha_2 - \gamma_2)(1 - \gamma_1)}.$$

Then, we fix the profile of strategies $\alpha_2, \gamma_2, \alpha_1$ and obtain the probabilities of cooperative choice for pure strategies $\gamma_1 = 0$ and $\gamma_1 = 1$.

For $\gamma_1 = 0$:

$$p_1^c = \frac{\alpha_1 - \alpha_2\alpha_1}{1 - \alpha_1(\alpha_2 - \gamma_2)},$$

$$p_2^c = \frac{\alpha_2 - \alpha_2\alpha_1 + \alpha_1\gamma_2}{1 - \alpha_1(\alpha_2 - \gamma_2)}.$$

When $\gamma_1 = 1$:

$$p_1^c = \frac{\alpha_1 + \alpha_2 - \alpha_2\alpha_1}{1 - (\alpha_2 - \gamma_2)(\alpha_1 - 1)},$$

$$p_2^c = \frac{\alpha_2 - \alpha_2\alpha_1 + \alpha_1\gamma_2}{1 - (\alpha_2 - \gamma_2)(\alpha_1 - 1)}.$$

In the symmetrical case, we have $\alpha_1 = \alpha_2, \gamma_1 = \gamma_2, \lambda_1 = \lambda_2$. Then, the probabilities of cooperative choice transform to the following equations:

$$p_1^c|_{\alpha_1=0} = \frac{\alpha\gamma}{1 + \gamma(\alpha - \gamma)}, \quad p_2^c|_{\alpha_1=0} = \frac{\alpha}{1 + \gamma(\alpha - \gamma)},$$

$$p_1^c|_{\alpha_1=1} = \frac{1 - \alpha + \alpha\gamma}{1 - (1 - \gamma)(\alpha - \gamma)}, \quad p_2^c|_{\alpha_1=1} = \frac{\gamma}{1 - (1 - \gamma)(\alpha - \gamma)},$$

$$p_1^c|_{\gamma_1=0} = \frac{\alpha - \alpha^2}{1 - \alpha(\alpha - \gamma)}, \quad p_2^c|_{\gamma_1=0} = \frac{\alpha - \alpha^2 + \alpha\gamma}{1 - \alpha(\alpha - \gamma)},$$

$$p_1^c|_{\gamma_1=1} = \frac{2\alpha - \alpha^2}{1 - (\alpha - 1)(\alpha - \gamma)}, \quad p_2^c|_{\gamma_1=1} = \frac{\alpha - \alpha^2 + \alpha\gamma}{1 - (\alpha - 1)(\alpha - \gamma)}.$$

Note that if we first made the symmetrical assumption before substituting pure strategies into resulted expressions, we would only obtain extreme cases.

Next, we find the payoff functions using (3) for particular strategies:

$$U|_{\alpha=0} = \frac{-4\alpha^2\gamma}{(\gamma(\alpha - \gamma) + 1)^2} - \frac{\alpha\gamma}{(\gamma(\alpha - \gamma) + 1)} + \frac{9\alpha}{(\gamma(\alpha - \gamma) + 1)} + 1,$$

$$U|_{\alpha=1} = \frac{-4\gamma(-\alpha(1 - \gamma) + 1)}{(-(\alpha - \gamma)(1 - \gamma) + 1)^2} + \frac{9\gamma}{(-(\alpha - \gamma)(1 - \gamma) + 1)} - \frac{(-\alpha(1 - \gamma) + 1)}{(-(\alpha - \gamma)(1 - \gamma) + 1)} + 1,$$

$$U|_{\gamma=0} = \frac{-(\alpha^2 + \alpha)}{(-\alpha(\alpha - \gamma) + 1)} - \frac{4(-\alpha^2 + \alpha)(-\alpha(\alpha - \gamma) + \alpha)}{(-\alpha(\alpha - \gamma) + 1)^2} + \frac{9(-\alpha(\alpha - \gamma) + \alpha)}{(-\alpha(\alpha - \gamma) + 1)} + 1,$$

$$U|_{\gamma=1} = \frac{-(\alpha(\alpha - 1) + \alpha)}{(-(\alpha - 1)(\alpha - \gamma) + 1)} - \frac{4(-\alpha(\alpha - 1) + \alpha)(-\alpha(\alpha - \gamma) + \alpha)}{(-(\alpha - 1)(\alpha - \gamma) + 1)^2}$$

$$+ \frac{9(-\alpha(\alpha - \gamma) + \alpha)}{(-(\alpha - 1)(\alpha - \gamma) + 1)} + 1.$$

Finally, we find QRE for PD in Markov strategies solving system of equations:

$$\begin{cases} \alpha = \dfrac{e^{\lambda U|_{\alpha=1}}}{e^{\lambda U|_{\alpha=0}} + e^{\lambda U|_{\alpha=1}}}, \\ \gamma = \dfrac{e^{\lambda U|_{\gamma=1}}}{e^{\lambda U|_{\gamma=0}} + e^{\lambda U|_{\gamma=1}}}, \end{cases}$$

where $\alpha \in [0; 1]$ and $\gamma \in [0; 1]$ are unknowns, and $\lambda \in [0; +\infty)$ is fixed. Note that different values of the rationality may lead to different profiles of strategies.

**Appendix 2**

Participants were recruited from the Moscow Institute of Physics and Technology in Moscow. A total of 168 individuals (72 females) participated in 14 experiments. We recruited participants through posting advertisements on the social networking site VKontakte. For every experiment, we only selected participants who were unacquainted with each other. Because our participants were students, we collected demographic information, such as academic major, group, and year of study. All participants were provided with written and verbal instructions related to the experiment. Experimenters notified participants that all points won in the games would be converted to real money (the average win rate was approximately equal to the cost of a full lunch in a cafe). The procedures of the study involving human participants were approved by the Tomsk State University Human Subjects Committee. Written informed consents were obtained from participants.

**Game.** The study employed the Prisoner's Dilemma Game (PD).

**Iterated Prisoner's Dilemma Game.** Two individuals anonymously participated in each round of the game. They both had two strategies: cooperation or defection (Table 1). Participants were divided into pairs randomly each period of the game.

**Experimental design.** The experimental procedure consisted of three stages. To execute the game, z-Tree, a specialized tool developed at the University of Zurich for designing and performing experiments in a group of experimental economics, was used[35].

**Stage 1: Anonymous playing phase.** Participants played the Prisoner's Dilemma Game for eleven to twenty-two game rounds. Participants did not know how many rounds they would play. In each round, participants were randomly divided into pairs and made choices simultaneously and independently of each other. In each round, participants were re-paired randomly, and participants were unaware of who they were playing against. After each of the periods, each participant observed their own and their opponent's results on a screen.

Points earned at this stage were added to the total win and converted into real money at the end of the game.

**Stage 2: Socialization phase.** In this phase, participants engaged in social interaction, which consisted of familiarization, communication, and division into groups. The participants memorized each other's names by playing "snowball"[9]. According to the game, players were seated in a circle, and the first person said his/her name and a personal quality that started with the same letter as the name. Second, the next participant repeated the name along with the quality of the first participant and gave his/her name and quality. This process was repeated with each participant until the last person, who was due to repeat all the names and personal qualities. Then, in a different order, the participants shared personal information, such as their hometown, academic major, hobbies, and interests. Following that, two captains were volunteer selected themselves from among the participants. The captains remained indoors while the other participants left the room. Then, in random order, they entered the room one by one. Every participant who entered the room chose a captain whose group he/she wanted to join. Consequently, two groups of 6 people were formed. In the end, each group of 6 people was tasked to find 5 common characteristics (i.e., 5 characteristics that united them) and choose a name for their group.

**Stage 3. Socialized phase.** The participants played the Prisoner's Dilemma Game. However, unlike the first stage, the participants interacted only in the groups of 6 previously composed during the socialization phase of the experiment. For each round, the participants were randomly divided into pairs. They were informed that they were interacting with a member of their "own" group, but they did not know who exactly that person was. After each of the periods, each participant observed their own and their opponent's results on a screen.

Both the games consisted of 15-20 rounds. The group names, chosen by participants at the socialization stage, appeared on monitors in the Prisoner's Dilemma Game.

Points were added to those obtained at the first stage. As a result, the final prize was generated and could be converted into a cash reward to compensate participants.

The schematic representation of the experiment design included the following components:

1. 12 participants are recruited (12 strangers).

2. Participants play PD in a mixed-gender group of 12 people for 11-22 rounds.

3. Socialization of unacquainted members of groups occurs. Participants are divided into two groups of 6 people.

4. Participants play PD in the newly formed groups for 15-20 rounds.

5. Participants are compensated for the experiment.

# Appendix 3

Table 1. Experimental results aggregated by experiments.

| Number of the experiment | % of cooperation before socialization | alpha before socialization | gamma before socialization | % of cooperation after socialization | alpha after socialization | gamma after socialization |
|---|---|---|---|---|---|---|
| Exp_1 | 18.89% | 0.19 | 0.22 | 36.67% | 0.34 | 0.43 |
| Exp_2 | 13.89% | 0.16 | 0.04 | 3875% | 0.34 | 0.45 |
| Exp_3 | 17.22% | 0.13 | 0.32 | 59.58% | 0.38 | 0.72 |
| Exp_4 | 29.17% | 0.25 | 0.36 | 88.33% | 0.69 | 0.91 |
| Exp_5 | 25.38% | 0.20 | 0.36 | 47.50% | 0.36 | 0.58 |
| Exp_6 | 14.02% | 0.12 | 0.14 | 53.75% | 0.23 | 0.80 |
| Exp_7 | 26.52% | 0.21 | 0.38 | 85.00% | 0.71 | 0.89 |
| Exp_8 | 9.47% | 0.07 | 0.17 | 32.92% | 0.26 | 0.43 |
| Exp_9 | 20.56% | 0.17 | 0.25 | 42.92% | 0.30 | 0.57 |
| Exp_10 | 28.33% | 0.31 | 0.19 | 35.42% | 0.27 | 0.49 |
| Exp_11 | 25.56% | 0.21 | 0.33 | 81.67% | 0.50 | 0.88 |
| Exp_12 | 20.45% | 0.18 | 0.29 | 67.50% | 0.37 | 0.81 |
| Exp_13 | 33.33% | 0.27 | 0.43 | 57.50% | 0.58 | 0.57 |
| Exp_14 | 28.79% | 0.27 | 0.25 | 84.44% | 0.73 | 0.86 |
| **Mean:** | **22.25%** | **0.20** | **0.27** | **58.00%** | **0.43** | **0.67** |